\documentstyle[11pt,aaspp4,flushrt]{article}
\def\drp {$\Delta m_{15}(B)$} 
\def\sig {$\sigma$}
\def\H {H$_0$}

\def\Hunits {km s$^{-1}$ Mpc$^{-1}$}
\def\kps {km s$^{-1}$}
\def\hiz {high--$z$}
\def\plotone#1{\centering \leavevmode
\epsfxsize=\textwidth \epsfbox{#1}}
\def\plottwo#1{\centering 
\epsfysize=8in \epsfbox{#1}}

\lefthead{Howell, Wang, \& Wheeler}
\righthead{The distribution of SNe Ia}

\begin{document}
\slugcomment{Accepted for publication in the Astrophysical Journal}
\begin{center}
\vspace{3.5mm}
\title{The Distribution of High and Low Redshift Type Ia Supernovae}
\author{D. Andrew Howell, Lifan Wang\altaffilmark{1}, and J. Craig Wheeler} 
\affil{Department of Astronomy and McDonald Observatory, University of Texas at Austin, Austin, TX 78712\\
howell@astro.as.utexas.edu, lifan@astro.as.utexas.edu, wheel@astro.as.utexas.edu}
\authoraddr{Astronomy Department, University of Texas, RLM 15.308, Austin, TX 78712-1083} 
\altaffiltext{1}{Hubble Fellow}
\end{center}

\begin{abstract}

The distribution of high redshift Type Ia supernovae (SNe Ia) 
with respect to projected distance from the center of the host 
galaxy is studied and compared to the distribution of local SNe.
The distribution of high--$z$ SNe Ia is found to be similar to the
local sample of SNe Ia discovered with CCDs, but different than the
sample discovered photographically.  This is shown to be due to the
Shaw effect.  These results have implications for the use of SNe Ia to 
determine cosmological parameters if the local sample of supernovae
used to calibrate the light curve decline relationships is drawn from
a sample discovered photographically.  A K--S test shows that the
probability that the high redshift SNe of the Supernova Cosmology Project
are drawn from the same distribution as the low redshift calibrators 
of Riess et al.\  is 0.1\%.  This is a potential problem because 
photographically discovered SNe are preferentially discovered farther
away from the galaxy nucleus, where SNe show a lower scatter in 
absolute magnitude, and are on average 0.3 magnitudes fainter than
SNe located closer to the center of their host galaxy.  This raises 
questions about whether or not the calibration SNe sample the full range of 
parameters potentially present in high redshift SNe Ia.  The limited
data available suggest that the calibration process is adequate;
however, it would be preferable if high redshift SNe and the low
redshift SNe used to calibrate them were drawn from the same sample,
as subtle differences may be important.  Data are 
also presented which suggest that the seeming anti-Malmquist trend noticed 
by Tammann et al.\ (1996, 1998) for SNe Ia in galaxies with Cepheid distances
may be due to the location of the SNe in their host galaxies.

\end{abstract}


\section{Introduction}
Statistical studies of SNe give us information about the Universe on all
scales.  They give us clues about the progenitors of SNe (impacting stellar
evolution), reveal the galactic SN rate (influencing galaxy evolution), and 
are being used to determine cosmological parameters (the evolution
of the Universe).  For all of these tasks, obtaining an unbiased sample of 
SNe, and thus fully understanding the selection effects involved in SN 
discovery, is essential.  This work will focus on selection effects 
that may affect the use of SNe Ia as distance indicators, so only SNe
Ia are discussed.

\subsection{SNe as Standard Candles}
All SNe Ia are thought to be caused by the thermonuclear 
explosions of carbon--oxygen (CO) white dwarfs in binary systems 
(Hoyle \& Fowler 1960).  
Their low dispersion in absolute magnitude allows them to be used as
``calibrated candles,''  though some care must be exercised if they
are to be used in this capacity.  The first Hubble diagram constructed
from Type I SNe by Kowal (1968) revealed a dispersion in photographic
magnitude of $\sim$0.6 mag.  Barbon et al.\ (1973) noted that Type I SNe were
not a homogeneous class~--- some light curves declined faster than others.
Pskovskii (1977,1984) then suggested that the light curve decline
rate is correlated with the luminosity of SNe.  Slower declining SNe Ia are
intrinsically more luminous than those with fast-declining light curves.
The separation of the Type Ib (Elias et al.\ 1985) and Ic subclasses
(Wheeler \& Harkness 1986), and the shift to modern
CCD detectors allowed Phillips (1993) to establish a decline rate parameter,
\drp, the decline in blue magnitude from maximum after 15 days, and to 
conclusively show that it was correlated with the absolute magnitudes of 
SNe at maximum.  In a study of 29 SNe Ia, Hamuy et al.\ (1996) have shown
that correcting the absolute magnitudes at maximum in this way reduces 
the dispersion from \sig = 0.38, 0.26, 0.19 to \sig = 0.17, 0.14, 0.13 in 
the B, V, and I bands respectively.  Another approach, the Multicolor
Light Curve Shape (MLCS) technique (Riess, Press, \& Kirshner, 1996),
makes use of the entire shape of the light curve in various colors to
reduce the dispersion at maximum light to $\sigma \simeq 0.15$ magnitudes.
These results have given promise that SNe Ia may be used as calibrated
candles to determine large scale motions of the local group (Riess, Press, 
\& Kirshner, 1995), and the cosmological parameters \H, $\Omega$, and 
$\Lambda$ (Goobar \& Perlmutter 1995).  Several searches have already 
discovered a combined number of $\sim$ 100 high redshift SNe 
($0.3 \leq z \leq 1.0$) for this purpose (Riess et al.\ 1998, Perlmutter 
et al.\ 1999), with the remarkable suggestion that $\Lambda$ is
finite and positive. 

\subsection{The Shaw Effect}
Shaw (1979) first characterized a selection effect in supernova searches 
that had long been suspected.  On photographic plates, supernovae are less 
frequently discovered in the often overexposed central regions of 
distant galaxies.  Shaw estimated that $\sim$50\% of supernovae are lost
within the central 8 kpc of galaxies beyond 150 Mpc.  The effect is 
reduced for closer galaxies.  A correction factor must be introduced to
account for this effect when determining SN rates.  Bartunov et al.\ (1992) 
and Cappellaro et al.\ (1993, 1997) find that supernova searches experience 
the Shaw effect to varying degrees, so the correction factor for each search
can be different.

To quantify the Shaw effect, one must compare supernovae discovered
photographically to an ``unbiased'' sample.  This has been attempted with
varying degrees of success, but so far each comparison group has had
problems.  Out of necessity, Shaw (1979), in his own words, made 
the ``unwarranted'' assumption that supernovae discovered in galaxies closer 
than 33 Mpc were free from this effect.  Prior to this, the loss of SNe in
the inner regions of galaxies was estimated by extrapolation of the SN rate 
in the outer regions of galaxies to a central peak following the light 
distribution (Johnson \& MacLeod 1963; McCarthy, 1974; Barbon et al.\ 1975).
Bartunov et al.\ (1992) confirmed Shaw's suspicion that even the nearby sample 
is not free from bias.  Bartunov et al.\ and Cappellaro et al.\ used combined 
data from visual and CCD searches as the control group.  

\subsection{Differences in SN Properties with Galactocentric Distance}
As pointed out by Wang, H\"oflich \& Wheeler (1997, hereafter WHW), SN 
properties appear
to vary with projected galactocentric distance (PGD).  Using 
data from 40 local well-studied SNe Ia, WHW find that SNe located more than 
7.5 kpc from the centers of galaxies show 3--4 times lower scatter in maximum 
brightness than those projected within that radius.  (Note that the 
projected distance is the minimum distance the SN can be from the center
of the galaxy.  Many events at small PGD are actually located far 
from the center of the host galaxy.)
A reduction of scatter with PGD is also apparent in the 
$B_{\rm max}-V_{\rm max}$ colors of the SNe.  In the absence
of corrections for light curve shape, SNe at higher
projected galactocentric distance are a more homogeneous group and should
be better for use as distance indicators.  The reddest and 
bluest, brightest and dimmest SNe are located near the galactic center, 
so extinction alone cannot explain the higher scatter in this region.  
Indeed, Riess et al.\ (1999) find that scatter still remains in these
relationships when the SNe are corrected for extinction using the
MLCS technique.  The scatter is reduced after the SNe are corrected for
the light curve decline relationship, leaving no apparent trend with
PGD.

As suggested by WHW and confirmed by Riess et al.\ (1999), SNe in the 
outer regions of galaxies show systematic differences in luminosity
with respect to those with smaller projected separations from the host
center.  After correcting for extinction, Riess et al.\ find that 
SNe at distances of 10 kpc or more from the centers of their host 
galaxies are dimmer by about 0.3 mag than the mean of those (projected) 
inside 10 kpc.  
Riess et al.\ also note that this effect may be related to one pointed out
by Hamuy et al.\ (1996) and Branch et al.\ (1996) --- SNe in early-type 
hosts tend to be dimmer.  At high PGD, the 
sample is dominated by SNe in elliptical host galaxies.
Possible sources for the variation in SN properties with projected distance
(if intrinsic) include metallicity gradients within the galaxy or differences 
in progenitor systems between the disks and bulges.

\section{The sample}
Five groups of SNe Ia were studied in this work - the local sample 
discovered with CCDs,
the local SNe discovered photographically, the high redshift SNe discovered
by the Supernova Cosmology Project, the local SNe used to calibrate the 
brightness of the \hiz\ SNe, and eleven very local supernovae discovered in
galaxies with Cepheid distances that are used to tie the SN distance scale
to the Cepheid distance scale.

The IAU Circular ``List of Supernovae'' web 
page\footnote{http://cfa-www.harvard.edu/cfa/ps/lists/Supernovae.html} 
was used to determine the 
initial sample.  The Asiago Supernova Catalog (Barbon et al.\ 1989) was used 
to provide missing data and the recession velocities of the host
galaxies.  Any recession velocities not in the Asiago catalog were 
taken from the NASA/IPAC Extragalactic Database (NED)\footnote{http://nedwww.ipac.caltech.edu/}.  
The sample was divided into high and low redshift at $z=0.3$. 

The SNe were retained if they met the following criteria:
\begin{itemize}
\item Type Ia
\item Projected offset of SN from center of host galaxy is known
\item Recession velocity or distance to host galaxy is known
\item Method of discovery is known
\end{itemize}
Distances to galaxies with recession velocities greater than 3000 \kps\
were computed using \H = 65 \Hunits.  For galaxies under 3000 \kps\ distances
were taken from Tully (1985) and scaled to \H = 65.

The distances to high redshift SNe were calculated using the angular size
distance for the $\Lambda=0$ model:
$$D=\frac{2c}{H_0} 
\left(\frac{(2-\Omega+\Omega z)-(2-\Omega)\sqrt{1+\Omega z}}
{\Omega^2(1+z)^2}\right)$$
(Fukugita et al.\ 1992) assuming $H_0=65$, $\Omega=0.2$.

The method of discovery (photographic, visual, CCD) was not listed 
in any database, so this information was obtained directly from the 
IAU Circulars reporting the discovery of each SN.  

When SNe are very near to the center of the host galaxy, positions are 
sometimes reported as ``very close'' to the center, and no separation
in arcseconds is given.  Four supernovae fell into this category,
SNe 1994U, 1996am, 1997dg, and 1998ci.  All are in the sample of local
SNe Ia discovered with CCDs.
Removing these SNe from the sample would have biased the results,
so they were retained and assigned zero offset from the center of 
the galaxy.  Presumably these SNe were located so close to the center
of their host galaxy that the offset can be considered negligible.

The high redshift SNe used were primarily those discovered by the 
Supernova Cosmology Project.  Finding charts were downloaded from 
the group's website\footnote{http://www-supernova.lbl.gov}.  Because the 
charts were in postscript format, galactocentric separations had to 
be estimated visually, so each separation measurement has some
degree of uncertainty associated with it, especially for SNe close
to the cores of galaxies.  Several of the SNe do have galactocentric 
separations reported in the circulars.  Whenever possible, these numbers 
were used.

The smallest group consists of eleven SNe that have been discovered in 
galaxies with Cepheid distances.  These supernovae are important 
for the determination of the Hubble constant, as they set the 
absolute distance scale.

The final group of SNe studied was the group of 27 local SNe used 
by Riess et al.\ 1998 to calibrate the MLCS and \drp\ techniques.
The Supernova Cosmology Project used only 18 local calibrators,
nearly a subset of the Riess et al.\ sample, so are not studied 
separately here.  Both sets are drawn from the Cal\'an/Tololo 
SN survey (see Phillips et al.\ 1999), so they were discovered by the 
same methods, and should have nearly identical properties.  
This amounts to comparing the high redshift SNe (mostly) from one
team with the low redshift SNe of another team, but this is not
expected to be a problem.  The two groups use similar discovery 
techniques for the high redshift SNe and nearly identical SNe for
the low redshift calibrators, so the conclusions of this paper
should be applicable to both groups.  We have chosen the only data
available at high redshift, and the most statistically significant 
sample of low redshift calibrators to study in this paper, and these 
happen to be from different research groups.

\section{Results}
\begin{figure}[!htp]
\begin{center}
\plottwo{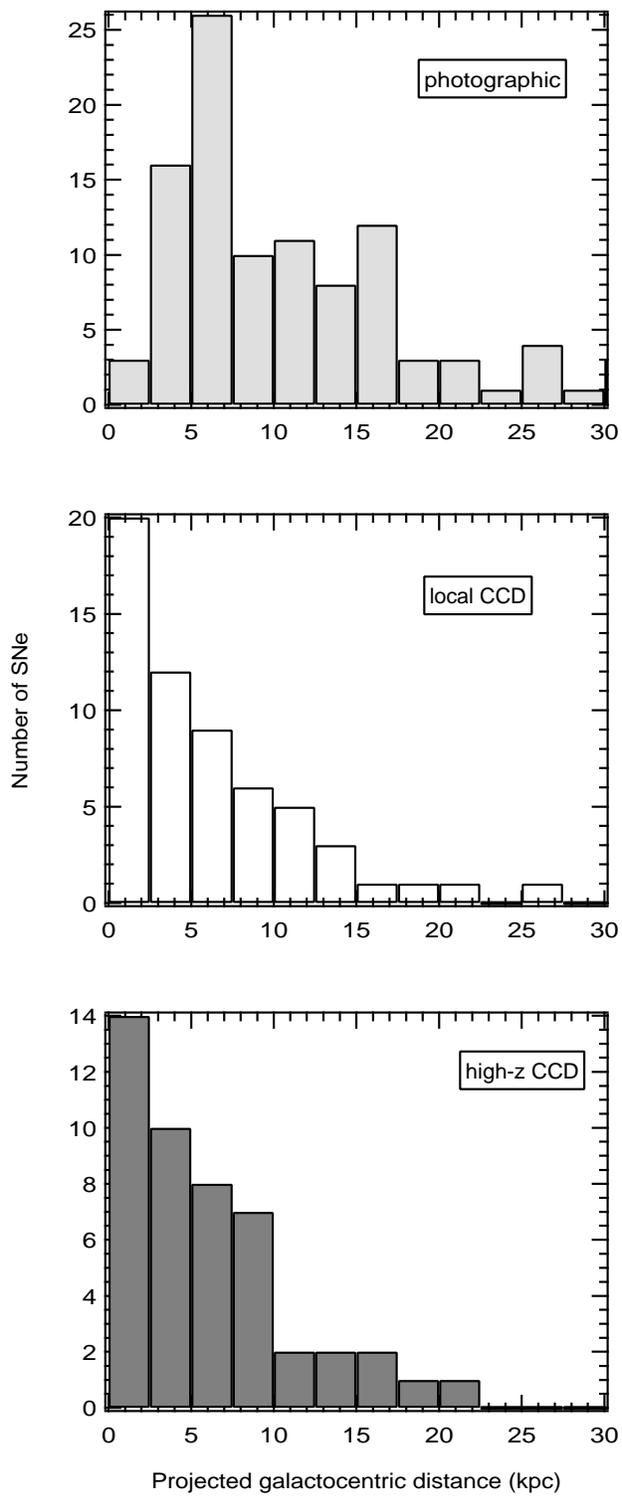}
\end{center}
\caption{The distribution of three samples of SNe Ia: those 
discovered photographically, local SNe discovered with CCDs, and 
\hiz\ SNe discovered with CCDs.}
\label{hist}
\end{figure}

Histograms of the distribution of SNe with respect to PGD were 
generated for SNe discovered with CCDs locally, at high redshift, and 
SNe discovered photographically.  These are presented in Fig.\ \ref{hist}.  
There are not enough data points in the Riess et al.\ calibration sample 
to generate a meaningful histogram.

SNe are presented with respect to raw PGD.  Care must be taken in the
interpretation of such data because it can be misleading and is 
impossible to properly normalize with the data available.  As one 
travels farther out in PGD, the area swept out by each histogram bin
increases.  Corrections for this effect are not posible
because there is no way to know the true distance of each SN from the
center of its host galaxy (only the projected distance is known).  
Also, the inclination is not known for each galaxy, and
the areal calibration would be different
for face-on spirals than it would be for edge-on spirals.  
It is noteworthy that SNe Ia discovered on CCDs are found in greater 
numbers in bins covering the smallest area --- those closer to the center
of the galaxy.  

Additional unremovable forms of bias are present in the data.  
The projection of a three dimensional galaxy onto the two dimensional sky
can cause interesting effects.  High PGDs are more likely to be
representative of true galactocentric distances.    
Consider an edge on spiral galaxy.  SNe at high PGD are confined to
a smaller range of distances perpendicular to the plane of the sky than
SNe observed near the center of the galaxy.  SNe with zero PGD may actually
have a very high true galactocentric distance.

The smaller size and faintness of galaxies at high redshift make them 
more susceptible to random errors in determining the PGD.  These random
errors can act like systematic errors because they tend to work in the
same direction.  SNe near the centers of galaxies may appear to have a larger 
PGD due to noise.  It is far more unlikely for noise to lower a PGD.

Human selection effects may also play a role in biasing the data.  SNe
near the centers of galaxies may be neglected at high redshift due to the
difficulty of obtaining spectra of SNe contaminated with galaxy light.

Each data set was sorted in order of increasing PGD and cumulative
distributions were generated, given in Figure \ref{ks}.  The left
axis can be interpreted as the fraction of SNe in each sample that
are projected interior to a given galactocentric distance.
The greatest vertical distance between two cumulative distribution 
curves is the Kolmogorov-Smirnov (K--S) statistic \emph{D} (Press et al.\ 
1988).  This is used to determine the K--S probability that two samples
were drawn from the same distribution.

A K--S test was done on each of the four data sets with respect to
each other and is presented in Table \ref{kstable}.  The null hypothesis
is that the two samples in each comparison were drawn from the same 
distribution.  The numbers in Table \ref{kstable} are the percent 
chance that the null hypothesis is true.  A few special cases 
deserve comment.

The probability that the sample of SNe discovered on photographic
plates is drawn from the same distribution as the sample of SNe
discovered on CCDs (at comparable redshifts, $z<0.3$) is $4 \times 10^{-5}$.  
As can be seen in Fig.\ \ref{ks}, the difference is due to the lack 
of SNe discovered near the centers of galaxies on photographic plates,
i.e. the Shaw Effect.  

In contrast, the K--S probability that local SNe discovered with CCDs and
\hiz\ SNe discovered with CCDs are drawn from the same distribution is
58\%.  This number is probably artificially low due to the uncertainties
in the positions of several local SNe very near the centers of galaxies
as mentioned above.  Indeed, when the sample is restricted to SNe
with PGD of 3 kpc or greater, the corresponding K--S probability increases 
to 74\%.  Note that according to K--S statistics, these probabilities 
approaching unity do not mean the distributions are the same, but that they
cannot be differentiated.

Figures \ref{hist} and \ref{ks} illustrate another selection effect caused 
by the differences between CCD and photographic discovery of SNe.  SNe 
in the outer regions of galaxies that are discovered by photographic 
means tend to be missed or omitted in CCD searches.  This is usually 
assumed to be 
due to the fact that CCDs have a smaller field of view and may not sample 
the outer regions
of galaxies.  One would thus expect that \hiz\ SNe would not show this effect, 
because their host galaxies have a smaller apparent angular diameter.  
Figure~\ref{ks} shows that the \hiz\ SNe presented in this study do show such 
a cutoff.  This is probably due to the limited number of data points.
Assuming SNe Ia arise from an exponential disk population with
a scale length of 5.0 kpc, the probablility of discovering a SN farther than
21 kpc from the center of the galaxy is 1.5\%.  In a sample of 49
SNe, it is not surprising that no SNe are seen at a galactocentric distance
greater than 21 kpc.  In addition, at high redshift, SNe at large PGD may 
not be followed up (Clocchiatti, private communication).

The sample of local SNe used to calibrate the \hiz\ SNe of Riess et al.\ 
(1998) was compared to the other groups of SNe.  Twenty-four of 
twenty-seven of 
these local calibrators (89\%) were discovered photographically, and
are also members of the sample of photographically discovered SNe studied
in this paper.
Figure \ref{ks} shows that the cumulative distribution of the local 
calibrators used by Riess et al.\  matches one expected from a 
photographic sample.  The K--S test gives the probability that the 
photographic sample and the Riess et al.\ calibrators are drawn from the
same distribution as 94\%.  This number is given only for completeness,
as the K--S test is not meaningful for significantly overlapping samples.

The most striking result is that the K--S test gives a probability of only 
$10^{-3}$ that the \hiz\ SNe are drawn from the same distribution
as the local SNe used to calibrate them.  The calibrators tend to be
located farther from the centers of galaxies than the \hiz\ sample,
so again the Shaw effect is largely to blame.  The implications for
the use of these SNe as standard candles are discussed below.

Tables \ref{vlocaltable} and \ref{mediantable} reveal another subtle
selection bias brought about by the Shaw Effect that has previously been
overlooked.  It is well known that SN discoveries in nearby galaxies are 
relatively unaffected by the Shaw Effect due to their large angular
size.  Of the eleven SNe with Cepheid distances, ten were located
at a PGD of less than 10 kpc (see Table \ref{vlocaltable}).  This is
particularly troubling given the findings of WHW and Riess et al.\ (1999) 
that SNe within this distance are on average 0.3 magnitudes brighter than
supernovae located farther from the centers of their host galaxies.
{\it The SNe that tie the distant SNe to the Cepheid distance scale are
drawn from a sample of potentially overluminous supernovae.}  This has 
implications for the determination of \H\, as discussed in the next section.

In summary, the closest SNe (which are used to tie all other SNe to the
Cepheid distance scale) are free from the Shaw Effect, but biased toward
overluminous SNe.  The intermediate SNe were largely discovered 
photographically, and are quite susceptible to the Shaw Effect.  As a
result, these supernovae (which are used as calibrators to train the 
MLCS technique and derive the light curve--decline relationship) are
biased toward underluminous SNe.  Finally, the \hiz\ SNe were discovered
with CCDs and are relatively immune to the Shaw Effect.  These results can 
easily be seen in Table \ref{mediantable}.  The median PGD for the photographic
sample is 9.1 kpc, compared to a median of only 4.4 kpc for the CCD sample.
The very local Cepheid calibrators have the smallest median PGD at only
3.2 kpc.

\section{Conclusions}

Recently Hatano et al.\ (1998) proposed an alternative explanation for the
Shaw Effect.  They speculated that the Shaw Effect in Type
II and Ibc SNe arises because extinction causes these supernovae to appear
dim when projected onto the centers of galaxies, and dimmer SNe are harder
to detect with increasing distance.  Their model does not predict a
paucity of SNe Ia near the centers of galaxies, and they were unconvinced 
that there is a real observational deficit of SNe Ia in this region.  The
data presented here confirm that the Shaw effect does indeed exist for 
photographically discovered Type Ia supernovae.  These data support 
the traditional interpretation of the Shaw Effect --- SNe are missed
due to saturation of galaxy cores on photographic plates.

\begin{figure}[!htp]
\begin{center}
\plotone{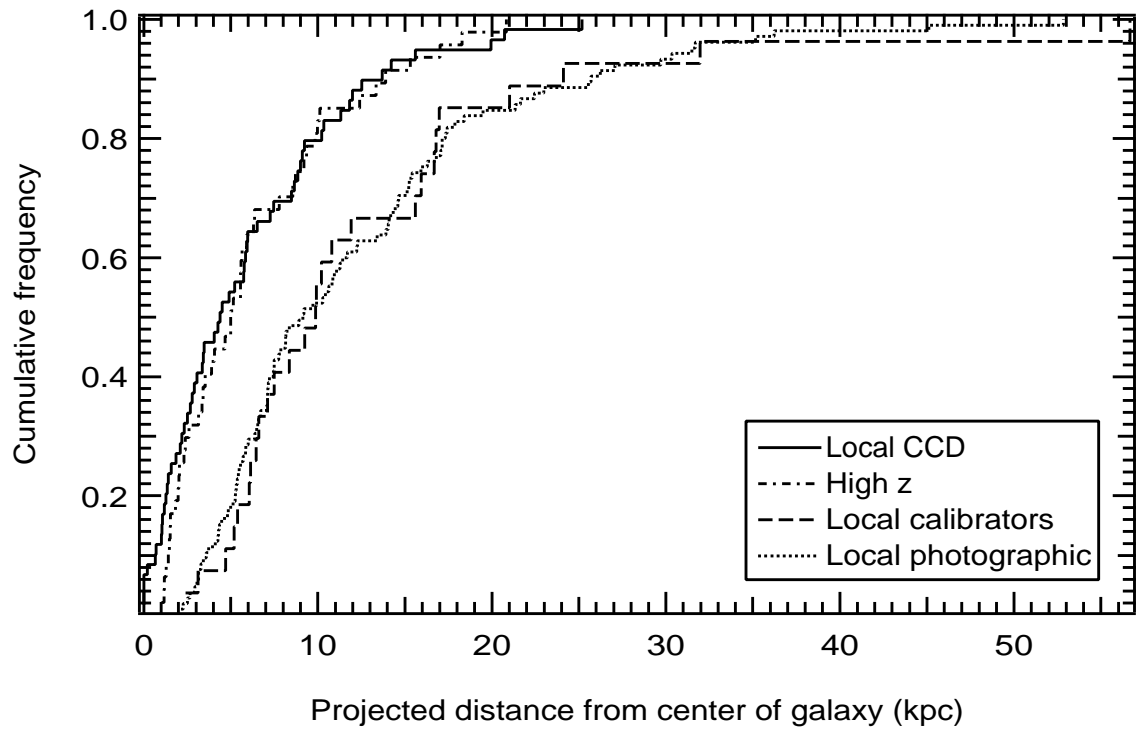}
\end{center}
\caption[The cumulative frequency of SNe Ia as a function of projected separation from 
the galactic nucleus]{The cumulative frequency of SNe Ia versus projected 
separation from the galactic nucleus.}
\label{ks}
\end{figure}

These data provide an estimate for the magnitude of the Shaw Effect
in SNe Ia.  If we assume the CCD sample is free from bias (the smaller
FOV bias is small as noted earlier), then 80\% of SNe should be located
within a PGD of 10 kpc (see Figure \ref{ks} and Table \ref{mediantable}).  
In the photographic
sample, 50 SNe have a PGD $\ge$ 10 kpc.  If this represents 20\% of the
sample, then the total sample should be $\sim$ 250 SNe;  however, there
are only 105 SNe in the sample, so $\sim$ 145 (58\%) are missing.
Thus nearly 60\% of SNe Ia are lost near the centers of galaxies on
photographic plates due to the Shaw Effect.  This is striking because
it is higher than most estimates of the Shaw Effect for all SNe, despite
the fact that SNe Ia are $\sim$ 1.5 times brighter on average than SNe II.   
This is the key piece of evidence that saturation (not extinction)
is to blame for the paucity of SNe.  It would be very interesting to do
a relative comparison of the samples of Type II and Type b/c SNe discovered
with CCD's compared to Type Ia SNe to explore the relative concentrations
toward the centers of galaxies.

The high redshift SN sample appears to be relatively free from selection bias
in terms of separation from the center of the galaxy.
These SNe were discovered with CCDs and show a distribution with respect
to galactocentric distance similar to that of local SNe discovered with
CCDs.  We find no evidence that \hiz\ SNe are selectively discovered farther 
from the centers of galaxies.  

The \hiz\ SN studies may be sensitive to selection effects.  According to
Riess et al.\ (1998), ``we must continue to be wary of subtle selection 
effects that might bias the comparison of SNe Ia near and far.''  They 
also add, ``It is unclear whether a photographic search selects
SNe Ia with different parameters or environments than a CCD search or
how this could affect a comparison of samples.  Future work on quantifying
the selection criteria of the samples is needed.''  

This raises a potential area of concern when using SNe Ia as distance 
indicators.  The light curve decline -- absolute magnitude
relationship used to calibrate the brightness of \hiz\ SNe is derived 
mainly from local SNe discovered photographically, as is the case in 
Riess et al.\ (1998) and Perlmutter et al.\ (1999). These SNe tend to be 
located farther from the centers of galaxies than the SNe in the \hiz\ 
sample.  The calibrators are thus drawn from a sample subject to selection 
bias that is both more uniform in luminosity and dimmer than average. 
Because the calibrators are likely more homogeneous as a group than the 
\hiz\ SNe, they may not be able to  effectively correct for the light 
curve decline relationship over all of parameter space.  

This work has shown that there are issues of selection bias to be considered,
but how much these issues affect the use of SNe Ia as distance indicators
remains unclear.  Fig.\ 6c of Riess et al.\ (1999) shows that after SN
magnitudes are corrected for light curve shape using the MLCS technique 
there is no apparent trend of absolute V magnitude at maximumn with PGD.
In this case the calibration process seems to work adequately.  No data is
available to indicate how well other calibration methods, particularly \drp, 
(used by Riess et al.\ 1998 in addition to MLCS) and the stretch factor 
method (used by 
Perlmutter et al.\ 1999), can compensate for changes in the absolute
magnitudes of SNe with PGD.  Since these calibration methods all rely on 
the same principle --- using the shape of the light curve to correct 
for variations in luminosity --- one might expect that all will work
similarly.  It is also not known how the calibration process affects
the trend reported by WHW of increased scatter in the colors of SNe
at maximum light with decreasing PGD.  Future work is needed to 
assure that all methods of calibration of SNe Ia can remove trends
with PGD in all bandpasses.  No previous studies have found significant
problems with the calibration process, so the results of Riess et al.\ (1998),
and Perlmutter et al.\ (1999) do not appear to be at risk due to 
selection bias.  Evidence for a local void (Zehavi et al.\ 1998),
already marginal, may be more sensitive to subtle effects.

The PGD-magnitude effect when combined with the Shaw effect may explain
several unusual findings reported recently.  Tammann et al.\ (1996) and
Tammann (1998) noted that after SNe luminosities are corrected via 
a light curve decline relationship, local SNe with Cepheid distances are 
brighter than average, and this is the opposite than one would expect from a 
magnitude limited sample (the farthest SNe should be brightest because we could
not see the faint ones).  This discrepency has been used to argue that
there is a problem with light curve decline relationships.  From the 
result presented here -- that ten of eleven SNe with Cepheid distances 
have PGD $<10$ kpc -- these SNe can be expected to be overluminous.  

Two of the SNe with reported Cepheid distances, 1980N and 1992A, do not 
actually have a Cepheid distance
determined to their host galaxies, but are members of the Fornax cluster,
for which a Cepheid distance of $(m-M)_0=31.35$ is available by assuming 
that the galaxy NGC 1365 is a member of the cluster (Madore et al.\ 1999).  
These two SNe appear to be too dim, by 0.3 mag, compared to the Cal\'an/Tololo
sample of SNe Ia, leading Suntzeff et al.\ (1999)
to speculate that NGC 1365 may be foreground to Fornax by 0.3 mag.
Interestingly, these SNe have the highest PGD of the SNe in Cepheid 
galaxies, at 20.0 and 9.3 kpc, respectively, both outside the distance
WHW identify as the cutoff beyond which SNe Ia are seen to be
0.3 mag dimmer on average.  A possible explanation for the faintness of
SN 1980N and SN 1992A
is their location in their host galaxies, rather than an error in
the distance to Fornax.   It should also be noted, however, that
Saha et al.\ (1999) derive a distance modulus to NGC 1316 of 
$(m-M)_0=31.84$, by assuming that SN 1980N in NGC 1316 is similar
to SN 1989B, which would make SN 1980N too bright.

The fact that the Cepheid calibrators are drawn from a potentially 
overluminous sample of SNe is troubling.  This could affect determinations 
of the Hubble constant
that rely on using these very local SNe for calibration purposes.  The
fact that different biases operate on different distance scales calls
into question the variation of \H\ with distance reported by Tammann (1998),
as no second parameter correction was used. 

More data is needed on the galactic radial distribution of high and low
redshift SNe.  Given the discovery rate of \hiz\ SNe and the fact that
they presumably do not face the discovery bias related to the small size
of a CCD relative to the host galaxy, the statistics of the intrinsic 
distribution of \hiz\ SNe may soon be known more precisely than those of 
the low redshift sample.

In an ideal world, the luminosity of \hiz\ SNe would only be calibrated
with SNe discovered using CCDs.  That is not possible at the present time,
so early work indicating that light curve decline -- absolute magnitude 
relationships can effectively remove the effects of selection bias should
be continued and strengthened.

\acknowledgements
The authors thank the anonymous referee for his or her helpful comments and
Adam Riess and Alan Sandage for comments and perspective.

This research was supported in part by NSF Grant 95-28110,
a grant from the Texas Advanced Research Program, and by
NASA through grant HF-01085.01-96A from the Space Telescope
Science Institute which is operated by the Association of
Universities for Research in Astronomy, Inc., under NASA
contract NAS 5-26555.

\begin{deluxetable}{lccc}
\tablewidth{0pt}
\tablecaption{K--S Probabilities \label{kstable}}
\tablehead{
\colhead{} &
\colhead{Calib} &\colhead{Hi-z}&\colhead{CCD} }
\startdata
Photo&94&0.008&0.004 \nl 
Calib&\nodata&0.1&0.05 \nl 
Hi-z&\nodata&\nodata&58 \nl
\enddata
\tablenotetext{}{
\,\newline
K--S percent probabilities that two samples were drawn from the same distribution.\\
Calib ---- The local SNe Ia used to calibrate the \drp\ and MLCS methods from Riess et al.\\
Photo --- SNe Ia discovered photgraphically between 1989 and 1998\\
Hi-z --- SNe Ia with $z>0.3$ discovered by the Supernova Cosmology Project\\
CCD --- Local SNe Ia discovered on CCDs between 1989 and 1998\\
}
\end{deluxetable}

\begin{deluxetable}{lccc}
\tablewidth{0pt}
\tablecaption{Projected Galactocentric Distances of Cepheid Calibrators \label{vlocaltable}}
\tablehead{
\colhead{SN}&\colhead{Host}&\colhead{D\tablenotemark{a}}&\colhead{PGD\tablenotemark{b}} \\
&&\colhead{(Mpc)} &\colhead{(kpc)}} 
\startdata
1895B&NGC 5253&4.3&0.7 \nl
1972E&NGC 5253&4.3&2.5 \nl
1937C&IC 4182&5.4&1.3 \nl
1981B&NGC 4536&16.6&4.1 \nl
1960F&NGC 4496&16.8&3.0 \nl
1990N&NGC 4639&25.5&8.0 \nl
1989B&NGC3627&11.4&2.8 \nl
1998bu&NGC 3368&11.9&3.2 \nl
1974G&NGC 4414&19.14&5.8 \nl
1992A&NGC 1380&18.6\tablenotemark{c}&9.3 \nl
1980N&NGC 1316&18.6\tablenotemark{c}&20.0 \nl
\enddata
\tablenotetext{a}{Distance to the host galaxy in Mpc (derived from data in Saha et al.\ (1999), Madore et al.\ (1999), Suntzeff et al.\ (1999), and references therein)}
\tablenotetext{b}{Projected Galactocentric Distance in kpc}
\tablenotetext{c}{Distance is actually to NGC 1365, assumed to also be a member of the Fornax cluster.  See discussion in text.}
\end{deluxetable}

\begin{deluxetable}{lcccr}
\tablewidth{0pt}
\tablecaption{Median PGD and SNe within 10 kpc \label{mediantable}}
\tablehead{
\colhead{Sample\tablenotemark{a}} & \colhead{Median\tablenotemark{b}}&
\colhead{N$_{PGD<10}$\tablenotemark{c}} & \colhead{N$_{tot}$\tablenotemark{d}} &\colhead{\%$_{PGD<10}$\tablenotemark{e}}\\
&\colhead{(kpc)} &&& }
\startdata
Local CCD          &4.4&47&59 &80\%\nl
Local Photo        &9.1&55&105&52\%\nl
High--z             &4.6&41&47&87\%\nl
Riess Calibrators  &9.9&15&27&56\%\nl
Cepheid Calibrators&3.2&10&11&91\%\nl
\enddata
\tablenotetext{a}{SN sample --- same as in Table 1, with the addition of
the Cepheid calibrators (SNe Ia with Cepheid distances to the host galaxy).}
\tablenotetext{b}{Median Projected Galactocentric Distance (PGD) --- distance between host galxy center and SN.}
\tablenotetext{c}{Number of SNe with PGD $<$ 10 kpc.}
\tablenotetext{d}{Total Number of SNe in the sample.}
\tablenotetext{e}{Percentage of SNe with PGD $<$ 10 kpc.}
\end{deluxetable}

\onecolumn

\newpage

\end{document}